\documentclass[twocolumn,secnumarabic,amssymb, amsmath, nofootinbib,tightenlines,
nobibnotes, aps, prl,epsfig]{revtex4}
\usepackage{graphicx}
\usepackage{dcolumn}
\usepackage{bm}
\begin{document}
\preprint{APS/123-QED}
\title{Dynamical
behavior connection of the gluon distribution and \\the proton
structure
function at small-$x$ }

\author{G.R.Boroun}%
 \email{grboroun@gmail.com; boroun@razi.ac.ir }
\affiliation{ Physics Department, Razi University, Kermanshah
67149, Iran}
\date{\today}
\begin{abstract}
We make a critical study of the relationship between the singlet
structure function $F_{2}^{S}$ and the gluon distribution
$G(x,Q^{2})$ proposed in Refs.[4-8], which is frequently used to
extract the gluon distribution from the proton structure function.
We show that a simple relation is not generally valid in the
simplest state. We completed this relation by using a
laplace-transform method and hard-Pomeron behavior at LO and NLO
at small-$x$. Our study show that this relation is dependence to
the splitting functions and initial conditions at
$Q^{2}=Q^{2}_{0}$ and running coupling constant at NLO. The
resulting analytic expression allow us to predict the proton
structure function with respect to the gluon distributions and to
compare the results with H1 data and a QCD analysis fit.
Comparisons with other results are made and predictions for the
proposed best
approach are also provided.\\
\end{abstract}
 \pacs{***}
\maketitle
\subsection{1.Introduction}
The Dokshitzer-Gribov-Lipatov-Altarelli-Parisi (DGLAP) evolution
equations $[1-3]$ are fundamental tools to study the ${\ln}Q^{2}$
and $x$ evolutions of structure functions, where $x$ and $Q^{2}$
are Bjorken scaling and the square of the four-vector momentum
exchange in deep inelastic scattering (DIS) process respectively.
The measurements of the $F_{2}(x,Q^{2})$ structure functions by
DIS processes in the small-$x$ region have opened up a new era in
parton density measurements inside hadrons. The structure function
reflects the momentum distributions of partons in a nucleon. It is
also important to know the gluon distribution inside a hadron at
small-$x$ because gluons are expected to be dominant in this
region. On the other hand, the gluon distribution functions cannot
be measured directly through experiments. It is, therefore,
important to measure the gluon distribution $G(x,Q^{2})$ directly
using the proton structure function $F_{2}(x,Q^{2})$. This
expectation has led to an approximate phenomenological scheme, as
in the past two decades some authors [4-8] reported an ansatz
between the gluon distribution function and singlet structure
function. The commonly used relation is
\begin{equation}
G(x,Q^{2})=K(x)F_{2}^{S}(x,Q^{2}),
\end{equation}
where $K(x)$ is a parameter to be chosen from the experimental
data and those assumed $K(x)=k, ax^{b}$ or $ce^{dx}$ where
$k,b,a,c$ and $d$ are constants. Authors used a Taylor expansion
for the gluon and singlet functions at low-$x$ in solving DGLAP
evolution equations with applying Eq.1 to the distribution
functions. As, Eq.1
 is a relationship between singlet structure function and
gluon distribution function was proposed in order to facilitate
the extraction of the gluon density from the data.\\
In this paper we deduce the general relations between the proton
structure function and the gluon distribution function with
analytical methods at leading order (LO) and next-to-leading order
(NLO). However, a relation between singlet structure function and
gluon distribution function can be determined by simultaneous
solutions of coupled DGLAP evolution equations of singlet
structure functions and gluon distribution functions.  We
demonstrate here that the validity of this relation crucially
depends on the splitting functions at LO and running coupling
constant at NLO. We derive the master equation to extract the
relation between the gluon distribution and the proton structure
function, by using a Laplace-transform technique at LO and also a
hard Pomeron behavior for the gluon distribution up to
next-to-leading order (NLO). Our purpose here is to improve the
situation with an approximation equation at small-$x$ at LO and
NLO. Section $2$ outlines the theory and formalism while section
$3$ is devoted to results and discussions.\\

\subsection{2.Compact Formula}
The DGLAP evolution equations for the singlet quark structure
function and the gluon density have the forms
\begin{eqnarray}
\frac{d}{d lnQ^{2}} \left[\begin{array}{c}
 q(x,Q^{2}) \\
 g(x,Q^{2})
 \end{array}
 \right]=\left[
 \begin{array}{cc}
 P_{qq} & P_{qg}\\
 P_{gq} & P_{gg}
 \end{array}
 \right]{\otimes}
\left[\begin{array}{c}
 q(x,Q^{2}) \\
 g(x,Q^{2})
 \end{array}
 \right]
\end{eqnarray}
which emphasized that quark and gluon densities are coupled. The
convolution, defined as $P_{ij}{\otimes}
\emph{f}_{j}=\int_{x}^{1}\frac{dy}{y}P_{ij}(\frac{x}{y})\emph{f}_{j}(y,Q^{2})$,
express the possibility that a parton $i$ with momentum fraction
$x$ may originate from the branching of a parent parton $j$ of the
higher momentum fraction $y$ ($P_{ij}$ is the splitting
function).\\
The method of approximate determination a relation between the
gluon and structure function is based on the simplification of the
convolutions $P_{ij}{\otimes} \emph{f}_{j}$ by the Laplace
transforms [9-12] and other methods based on the behavior of the
gluon distribution such as the hard Pomeron and the expanding
methods [13-15]. Here we present a general solution of the DGLAP
evolution equations at low-$x$, as the gluons are expected to be
dominant. Therefore we can neglect the quark singlet part to the
evolution equations and also the non-singlet contribution
$F_{2}^{NS}$ can be ignored safely at small-$x$ in the DGLAP
equations. Complete solution of the decoupling DGLAP evolution
equations for a relation between gluon and
singlet functions can be discussed at section 2.2.\\
 The
LO DGLAP equations for the singlet and gluon functions can be
written as
\begin{eqnarray}
\frac{4\pi}{\alpha^{\rm
LO}_{s}(Q^{2})}\frac{{\partial}F^{S}_{2}(x,Q^{2})}{{\partial}{\ln}Q^{2}}&{\simeq}&2n_{f}x{\int_{x}^{1}}
G(z,Q^{2})(1-2\frac{x}{z}\nonumber\\
&&+2\frac{x^{2}}{z^{2}})\frac{dz}{z^{2}},
\end{eqnarray}
and
\begin{eqnarray}
\frac{4\pi}{\alpha^{\rm
LO}_{s}(Q^{2})}\frac{{\partial}G(x,Q^{2})}{{\partial}{\ln}Q^{2}}{\simeq}\frac{33-2n_{f}}{3}G(x,Q^{2})\\\nonumber
+12G(x,Q^{2})\ln\frac{1-x}{x}\\\nonumber
 +12x{\int_{x}^{1}}
(\frac{G(z,Q^{2})}{z}-\frac{G(x,Q^{2})}{x})\frac{dz}{z-x}\\\nonumber
+12x{\int_{x}^{1}}
G(z,Q^{2})(\frac{z}{x}-2+\frac{x}{z}-\frac{x^{2}}{z^{2}})\frac{dz}{z^{2}}.
\end{eqnarray}
Here $\alpha^{\rm LO}_{s}(Q^{2})$ is given by the LO form
\begin{equation}
\alpha_{s}^{\rm
LO}=\frac{4\pi}{(11-\frac{2}{3}n_{f}\ln(Q^{2}/\Lambda^{2}))},
\end{equation}
where $n_{f}$ being the number of active quark flavors($n_{f}=4$) and $\Lambda$ is the QCD cut-off parameter.\\

\subsection{2.1. Laplace Transform method}
Authors in Ref.[9-12] uses a somewhat unusual application
transforms, in which those first introduce the variable
\begin{eqnarray}
\upsilon &{\equiv}& \ln(1/x),
\end{eqnarray}
into the coupled DGLAP evolution equations, then obtained the
coupled equations in the Laplace-space variable $s$, as we can be
written these equations at small-$x$ for our limit
\begin{eqnarray}
\frac{{\partial}F_{2}^{S}}{{\partial}{\ln}Q^{2}}(s,Q^{2}){\simeq}\frac{\alpha^{\rm
LO}_{s}(Q^{2})}{4\pi}\Theta_{F}(s)G(s,Q^{2}),
\end{eqnarray}
and
\begin{eqnarray}
\frac{{\partial}G}{{\partial}{\ln}Q^{2}}(s,Q^{2}){\simeq}\frac{\alpha^{\rm
LO}_{s}(Q^{2})}{4\pi}\Phi_{G}(s)G(s,Q^{2}),
\end{eqnarray}
where $F(s)=
{\mathcal{L}}[\hat{F}(\upsilon);s]=\int_{0}^{\infty}\hat{F}(\upsilon)e^{-s\upsilon}d\upsilon
$ and $G(s)=
{\mathcal{L}}[\hat{G}(\upsilon);s]=\int_{0}^{\infty}\hat{G}(\upsilon)e^{-s\upsilon}d\upsilon
$ ($\hat{F}(\upsilon){\equiv}F(e^{-\upsilon}),
\hat{G}(\upsilon){\equiv}G(e^{-\upsilon})$). The coefficient
functions $\Phi_{G}(s)$ and $\Theta_{F}(s)$ are given by [9-12]
\begin{eqnarray}
\Theta_{F}(s)=2n_{f}(\frac{1}{1+s}-\frac{2}{2+s}+\frac{2}{3+s})
\end{eqnarray}
and
\begin{eqnarray}
\Phi_{G}(s)=\frac{33-2n_{f}}{3}+12(\frac{1}{s}-\frac{2}{1+s}-+\frac{1}{2+s}\\\nonumber
-\frac{1}{3+s}-\psi(1+s)-\gamma_{E})
\end{eqnarray}
where $\psi(x)$ is the digamma function and
$\gamma_{E}=0.5772156...$ is Euler$_{^{,}}$s constant.\\
For obtain an general explicit form between the gluon distribution
and the proton structure function at small-$x$,  rewrite Eqs.7 and
8 in $s$ space as
\begin{eqnarray}
\frac{{\partial}G}{{\partial}{\ln}Q^{2}}(s,Q^{2})=\frac{\Phi_{G}(s)}{\Theta_{F}(s)}\frac{{\partial}F_{2}^{S}}{{\partial}{\ln}Q^{2}}(s,Q^{2}).
\end{eqnarray}
or
\begin{eqnarray}
\frac{{\partial}G}{{\partial}{\ln}Q^{2}}(s,Q^{2})=h(s)\frac{{\partial}F_{2}^{S}}{{\partial}{\ln}Q^{2}}(s,Q^{2}).
\end{eqnarray}
In the above equation we used the following property for Laplace
transformation
\begin{eqnarray}
{\mathcal{L}}^{-1}[h(s)\frac{{\partial}F_{2}^{S}(s,Q^{2})}{{\partial}{\ln}Q^{2}};\upsilon]=\int_{0}^{\upsilon}\frac{{\partial}\widehat{F_{2}^{S}}(w,Q^{2})}{{\partial}{\ln}Q^{2}}
\widehat{H}(\upsilon-w)dw,\nonumber\\
\end{eqnarray}
where
$\widehat{H}(\upsilon){\equiv}{\mathcal{L}}^{-1}[h(s),\upsilon]$.
The calculation of $\widehat{H}(\upsilon)$, using Eqs.9 and 10,
for LO is straightforward and given by
\begin{eqnarray}
\widehat{H}(\upsilon)&=&\frac{9}{4}+\frac{13}{8}~\delta(\upsilon)+\frac{25}{24}~\delta'(\upsilon)\\\nonumber
&&-e^{(\frac{-3}{2}\upsilon)}[\frac{\sqrt{7}}{7}\sin(\frac{\sqrt{7}}{2}\upsilon)+\frac{13}{3}\cos(\frac{\sqrt{7}}{2}\upsilon)].
\end{eqnarray}
Here we neglecting the some terms at small-$x$, as
$12\int_{0}^{\upsilon}\frac{\partial \widehat{G}}{\partial
w}(w,Q^{2}) \ln(1-e^{-(^{\upsilon-w})})dw {\rightarrow}0$.
Therefore we obtain an explicit solution for the derivatives of
the gluon distribution in terms of the integral
\begin{eqnarray}
\frac{{\partial}\widehat{G}(\upsilon,Q^{2})}{{\partial}{\ln}Q^{2}}=\int_{0}^{\upsilon}\frac{{\partial}\widehat{F}_{2}^{S}(w,Q^{2})}{{\partial}{\ln}Q^{2}}
\widehat{H}(\upsilon-w)dw.\nonumber\\
\end{eqnarray}
Transforming back into $x$-space, finally we have an approximate
approach to the relation between the gluon distribution and
singlet structure function at low-$x$ by the following form
\begin{eqnarray}
G(x,Q^{2})&=&G(x,Q_{0}^{2})+\frac{13}{8}\mathcal{FF}(x)\nonumber\\
&&+\frac{9}{4}\int_{x}^{1}\mathcal{FF}(z)\frac{dz}{z}-\frac{25}{24}x\frac{{\partial}\mathcal{FF}(x)}{{\partial}x}\nonumber\\
&&-\int_{x}^{1}\mathcal{FF}(z)(\frac{x}{z})^{\frac{3}{2}}(\frac{\sqrt{7}}{7}\sin(\frac{\sqrt{7}}{2}\ln\frac{z}{x})\nonumber\\
&&+\frac{13}{3}\cos(\frac{\sqrt{7}}{2}\ln\frac{z}{x}))\frac{dz}{z},
\end{eqnarray}
 where
$\mathcal{FF}(x){\equiv}F^{S}_{2}(x,Q^{2})-F^{S}_{2}(x,Q_{0}^{2})$
and
$\mathcal{FF}(z){\equiv}F^{S}_{2}(z,Q^{2})-F^{S}_{2}(z,Q_{0}^{2})$.
Therefore the gluon distribution can be expressed into the singlet
structure function by Eq.16 with respect to the initial
conditions. This result is general with respect to the
approximated limit for the coupled DGLAP evolution equations at
small-$x$, and its the simplest answer to the relation between the
gluon distribution
and singlet structure function by using a Laplace-transform method.\\

\subsection{2.2. Hard-Pomeron behavior}
With respect to the Regge-like behavior of the gluon distribution
at small-$x$, we would like to get a simplest formulae to extract
the gluon distribution with respect to the proton singlet
structure function [13-15]. Authors in Refs.[16-17] shown a simple
relation between the gluon and $F_{2}$ at small-$x$ based on the
coupled integro-differential equations as can be converted in more
simple linear relations between the gluon distribution and
structure function and its derivatives with respect to
${\ln}Q^{2}$. The authors results  in Refs.[16-17] are different
from Eq.1 as it was proposed in the literature [4-8] to isolate
the
gluon distribution only with the singlet structure function.\\
The small-$x$ region of DIS offers a unique possibility to explore
the Regge limit of pQCD. This theory is successfully described by
the exchange of a particle with appropriate quantum numbers and
the exchange particle is called a Regge pole. Phenomenologically,
the Regge pole approach to DIS implies that the structure
functions are sums of powers in $x$, modulus logarithmic terms,
each with a $Q^{2}$- dependent residue factor. This model gives
the following parametrization of the DIS structure function
$F_{2}(x,Q^{2})$ at small x, $F_{2}(x,Q^{2})=A(Q^{2})x^{-\delta}$,
that the singlet part of the structure function is controlled by
Pomeron exchange at small $x$. The rapid rise in $Q^{2}$ of the
structure functions was considered as a sign of departure from the
standard Regge behavior. In principle, the HERA data should
determine the small-$x$ behavior of the gluon and sea-quark
distribution. Roughly speaking, the data on the singlet part of
the structure function $F_{2}$ constrain the sea quarks and the
data on the slope $dF_{2}/d\ln Q^{2}$ determine the gluon density.
In the DGLAP formalism, the gluon splitting functions are singular
as $x{\rightarrow}0$. Thus, the gluon distribution will become
large as $x{\rightarrow}0$, and its contribution to the evolution
of the parton distribution becomes dominant. In particular, the
gluon will drive the quark singlet distribution, and, hence, the
structure function $F_{2}$ becomes large as well, the rise
increasing in steepness as $Q^{2}$ increases [18-21]. Therefore,
the small $x$ limit corresponds to a study of a partonic system
inside of a nucleon which is predominantly formed by gluons. This
strong rise can eventually violate unitarity and so it has to be
tamed by screening effects. However when the density becomes large
enough, the gluons start interacting with each others and then
their further evolution is non-linear. This happens reduce the
growth of gluon distribution and called parton saturation [22-31].
Therefore, the linear evolution equation in this case is modified
by non-linear term description gluon recombination. An important
point in the gluon saturation approach is the $x$-dependent
saturation scale $Q_{s}^{2}(x)$. This scaling argument leads to
the conclusion that $\gamma^{*}p$ cross section, which is a priori
function of two independent variable ($x$ and $Q^{2}$), is a
function of only variable $\tau=\frac{Q^{2}}{Q_{s}^{2}(x)}$ where
the saturation scale is given by
$Q_{s}^{2}(x)=Q_{0}^{2}(\frac{x}{x_{0}})^{-\lambda}$ [32-34] and
its known as geometrical scaling. Here $Q_{0}$ and $x_{0}$ are
free parameters and exponent $\lambda$ is a dynamical quantity of
the order $\lambda{\sim}0.3$, although one can take into account
phenomenologically where exponent $\lambda$ has an effective
$Q^{2}$ dependence $\lambda=\lambda_{phn}(Q^{2})$. For
$Q^{2}<Q_{s}^{2}(x)$ such a scaling is natural, whereas for large
$Q^{2}>Q_{s}^{2}(x)$ it is a consequence of hard-poemron behavior
from hard diffraction [22-34].\\
At small $x$, $Q_{s}^{2}(x){\gg}\Lambda_{QCD}$ and the approach
based on PQCD is fully justified and results are based on the
phenomenon of geometric scaling. All results to DIS data from HERA
for $x<0.01$ show that geometrical scaling was found in the data
from different experiments. In the limit of high energy, PQCD
consistently predicts that the high gluon density should form a
Color Glass Condensate (CGC), where the interaction probability in
DIS becomes large and this is characterized by a hard saturation
scale $Q_{s}(x)$ which grows rapidity with $1/x$ [22-31]. In this
region, the nonlinear saturation dynamics is incorporated into the
CGC model. As, it is valid only for $Q^{2}$ less than or of the
order of the saturation momentum, which is at most several
$GeV^{2}$, while the fit result to SGK [25] model  extends up to
$Q^{2}$ of the order of several hundred $GeV^{2}$. Indeed, the
extended scaling at $Q^{2}>Q^{2}_{s}$ arises from the general
non-linear evolution equations in the kinematical range. The
validity of these evolution equations in the present of saturation
has been estimated as
$Q^{2}_{s}(x)<Q^{2}<Q^{4}_{s}(x)/\Lambda^{2}_{QCD}$. This means
that the geometric scaling for all momenta $Q^{2}$ have to
satisfied this inequality as, $\ln(Q^{2}/Q^{2}_{s}
){\ll}\ln(Q^{2}_{s}/\Lambda^{2}_{QCD})$. At soft momenta
$Q^{2}{\leq}Q^{2}_{s}$ this scaling is an expected consequence of
saturation and  at high momenta $1<\ln(Q^{2}/Q^{2}_{s}
){\ll}\ln(Q^{2}_{s}/\Lambda^{2}_{QCD})$ it rather corresponds to a
regime where parton densities are small, and linear evolution
equations apply.\\
 The overall physical picture is dependence to
the different regions in the ($x,Q^{2}$)-plane. For
$Q^{2}<Q_{s}^{2}(x)$ the linear evolution is strongly perturbed by
nonlinear effects where the parton system becomes dense and the
saturation corrections start to play an important role. In this
region the dipole cross section is bounded by an energy
independent value, as the dipole cross section was proposed
[22-31] to have the form
$\sigma_{dipole}(x,r)=\sigma_{0}\{1-exp(-r^{2}Q_{s}^{2}(x)/4)\}$
which impose the unitarity condition
($\sigma_{q\overline{q}}{\leq}\sigma_{0}$) for large dipole sizes
$r$. At small -$r$ region, the dipole cross section is related to
the gluon density where it is valid in the double logarithmic
approximation. This geometrical scaling holds until the line
boundary where $Q^{2}=Q_{s}^{2}(x)$. As  the gluon density is
$xg(x,Q^{2}=Q_{s}^{2}(x))=r^{0}x^{-\lambda}$, and the parameter
$r_{0}$ specifies the normalization along the critical line. Thus,
the saturation scale is an intrinsic characteristic of a dense
gluon system. For $Q^{2}>Q_{s}^{2}(x)$ the nonlinear screening
effects can be neglected and evolution of parton densities is
governed by the linear DGLAP equations [35-37]. Therefore, the
validity of our method only holds in the kinematic region
$Q^{2}{\gg}Q_{s}^{3}/\Lambda_{QCD}$. Hence, as $x$ gets smaller,
the gluon distribution grows rapidly and
$\lambda{\rightarrow}{\delta}$ where ${\delta}$
 is the hard pomeron exponent. So, the dipole cross section extracted from DIS data with assuming a hard pomeron dependence, as
 $\sigma {\sim}x^{-\delta}$. Therefore we study the DGLAP evolution upon the geometrical scaling in
 the region $Q^{2}>Q_{s}^{2}(x)$ with solving the linear DGLAP
 evolution equation starting from the gluon distribution
 satisfying the hard-pomeron behavior. The gluon distribution at small-$x$
increase with decreasing $x$ as
\begin{equation}
G(x,Q^{2})= f(Q^{2})x^{-\delta}.
\end{equation}
The form $x^{-\delta}$ of the gluon parameterization at small $x$
is suggested by Regge behavior, but because the conventional Regge
exchange is that of a soft Pomeron, with $\delta{\sim}0$, we may
also allow a hard Pomeron with $\delta{\sim}0.5$ [18-21]. Based on
the hard Pomeron behavior for the gluon distribution, let us put
Eq.(17) in Eqs.3 and 4. Let us introduce the variable
$y=\frac{x}{z}$. After doing the integration over $y$ , Eqs.3 and
4 can be rewritten as
\begin{eqnarray}
\frac{4\pi}{\alpha^{\rm
LO}_{s}(Q^{2})}\frac{{\partial}F^{S}_{2}(x,Q^{2})}{{\partial}{\ln}Q^{2}}&{\simeq}&2n_{f}G(x,Q^{2}){\times}\nonumber\\
&&{\int_{x}^{1}} z^{\delta}(1-2y+2y^{2})dy,
\end{eqnarray}
and
\begin{eqnarray}
\frac{4\pi}{\alpha^{\rm
LO}_{s}(Q^{2})}\frac{{\partial}G(x,Q^{2})}{{\partial}{\ln}Q^{2}}{\simeq}\frac{33-2n_{f}}{3}G(x,Q^{2})\\\nonumber
+12G(x,Q^{2})\ln\frac{1-x}{x}\\\nonumber
 +12G(x,Q^{2}){\int_{x}^{1}}
(y^{1+\delta}-1)\frac{dy}{y(1-y)}\\\nonumber
+12G(x,Q^{2}){\int_{x}^{1}} y^{\delta}({y}^{-1}-2+y-y^{2})dy.
\end{eqnarray}
Consequently
\begin{eqnarray}
\frac{{\partial}G(x,Q^{2})}{{\partial}{\ln}Q^{2}}=\frac{g1}{p1}\frac{{\partial}F^{S}_{2}(x,Q^{2})}{{\partial}{\ln}Q^{2}},
\end{eqnarray}
where
\begin{eqnarray}
p_{1}=2n_{f}{\int_{x}^{1}} z^{\delta}(1-2y+2y^{2})dy,
\end{eqnarray}
and
\begin{eqnarray}
g_{1}=\frac{33-2n_{f}}{3}+12\ln\frac{1-x}{x}+12{\int_{x}^{1}}
(y^{1+\delta}-1)\frac{dy}{y(1-y)}\nonumber\\
 +12{\int_{x}^{1}}
y^{\delta}(\frac{1}{y}-2+y-y^{2})dy.\nonumber\\
\end{eqnarray}
Eq.20 is independent of the running coupling constant
($\alpha_{s}(Q^{2})$) at LO. After successive integrations of both
sides of Eq.20, and some rearranging, we find an simplest equation
which determine $G(x,Q^{2})$ in terms of $F^{S}_{2}(x,Q^{2})$.
Consequently
\begin{eqnarray}
G(x,Q^{2})=\frac{g_{1}}{p_{1}}F^{S}_{2}(x,Q^{2})+[G(x,Q_{0}^{2})-\frac{g_{1}}{p_{1}}F^{S}_{2}(x,Q_{0}^{2})].\nonumber\\
\end{eqnarray}
 We observe that this equation demonstrates the close relation between $G(x,Q^{2})$ and $F^{S}_{2}(x,Q^{2})$ at small-$x$ into
 the initial conditions at $Q_{0}^{2}$ at LO by using a hard-Pomeron behavior for the gluon distribution.\\
The NLO corrections are add to LO, as the splitting functions
$P_{ij}^{,}s$ are the LO and NLO Altarelli- Parisi splitting
kernels by the following form
\begin{eqnarray}
P_{ij}(x,\alpha_{s}(Q^{2}))=P_{ij}^{\rm
LO}(x)+\frac{\alpha_{s}(Q^{2})}{2\pi}P_{ij}^{\rm NLO}(x).
\end{eqnarray}
The next-to-leading order is the standard approximation for most
important processes. The corresponding one- and two-loop splitting
functions have been known for a long time. Also, the NNLO
corrections can be need to be included, in order to obtain a
quantitatively reliable predictions for hard processes at present
and future high-energy colliders [38].\\
The running coupling constant $\frac{\alpha_{s}}{2\pi}$ has the
form in the LO and NLO respectively
\begin{equation}
\frac{\alpha_{s}^{\rm
LO}}{2\pi}=\frac{2}{\beta_{0}\ln\frac{Q^{2}}{\Lambda^{2}}},
\end{equation}
and
\begin{equation}
\frac{\alpha_{s}^{\rm
NLO}}{2\pi}=\frac{2}{\beta_{0}\ln\frac{Q^{2}}{\Lambda^{2}}}[1-\frac{\beta_{1}{\ln}\ln\frac{Q^{2}}{\Lambda^{2}}}{\beta_{0}^{2}\ln\frac{Q^{2}}{\Lambda^{2}}}],
\end{equation}
where $\beta_{0}=\frac{1}{3}(33-2N_{f})$ and
$\beta_{1}=102-\frac{38}{3}N_{f}$ are the one-loop and two-loop
corrections to the QCD $\beta$-function.\\
Therefore the DGLAP evolution equations have these behavior at NLO
 with respect to the hard-Pomeron behavior at small-$x$, as we have
 \begin{equation}
\frac{dG(x,Q^{2})}{d\ln Q^{2}}=\frac{\alpha_{s}}{4\pi}[g_{1}
+(\frac{\alpha_{s}}{4\pi})g_{2} ]G(x,Q^{2}),
\end{equation}
and
 \begin{equation}
\frac{dF_{2}^{S}(x,Q^{2})}{d\ln
Q^{2}}=\frac{\alpha_{s}}{4\pi}[p_{1}
+(\frac{\alpha_{s}}{4\pi})p_{2} ]G(x,Q^{2}),
\end{equation}
 where
\begin{eqnarray}
g_{2}&=&2\frac{(12C_{F}n_{f}T_{R}-46C_{A}n_{f}T_{R})}{9\delta}(1-x^{\delta})\nonumber\\
&&+2[n_{f}T_{R}(\frac{-61}{9}C_{F}+\frac{172}{72}C_{A})+C^{2}_{A}(\frac{1643}{54}-\frac{22}{3}\zeta(2)\nonumber\\
&&-8\zeta(3))]\frac{1-x^{1+\delta}}{1+\delta},
\end{eqnarray}
and
\begin{equation}
p_{2}=\frac{\alpha_{s}}{4\pi}\frac{80C_{A}N_{f}T_{R}}{9\delta}(1-x^{\delta}),
\end{equation}
where $p_{2}$ is the NLO kernel after doing the integration based
on the hard Pomeron behavior at Eq.28 according to the NLO
splitting function in Appendix. Therefore the close relation
between the gluon distribution and singlet structure functions at
NLO, when the coupling is fixed, is given by
\begin{equation}
G(x,Q^{2})=kF^{S}_{2}(x,Q^{2})+[G(x,Q_{0}^{2})-kF^{S}_{2}(x,Q_{0}^{2})],
\end{equation}
where
\begin{equation}
k=\frac{g_{1}+{\frac{\alpha_{s}}{4\pi}}g_{2}}{p_{1}+{\frac{\alpha_{s}}{4\pi}}p_{2}}.
\end{equation}
We now pass to the more realistic case with running coupling. In
this case the relation between the distribution functions takes
the form
\begin{equation}
G(x,Q^{2})=G(x,Q_{0}^{2})+\int_{Q_{0}^{2}}^{Q^{2}}k\frac{{\partial}F^{S}_{2}(x,Q^{2})}{{\partial}{\ln}Q^{2}}d{\ln}Q^{2}.
\end{equation}
Similarly, We get the singlet structure function evolution at NLO,
as
\begin{equation}
F^{S}_{2}(x,Q^{2})=F^{S}_{2}(x,Q_{0}^{2})+\int_{Q_{0}^{2}}^{Q^{2}}k'\frac{{\partial}G(x,Q^{2})}{{\partial}{\ln}Q^{2}}d{\ln}Q^{2},
\end{equation}
where
\begin{equation}
k'=\frac{p_{1}+{\frac{\alpha_{s}}{4\pi}}p_{2}}{g_{1}+{\frac{\alpha_{s}}{4\pi}}g_{2}}.
\end{equation}
The expansion of the results from NLO to NNLO approximation can be
done easily. Here we used our approximation approach to obtained a
simplest relation between the gluon distribution and singlet
structure function. The complete calculation of the DGLAP
evolution equations, when the singlet quark distribution is
essentially driven by the generic instability of the gluon
distribution, can be down numerically for shown that what is the
best relation between the distribution functions at LO up to NNLO.
In a resent paper [38] the distribution functions have been
obtained by solving decoupling DGLAP evolution equations at LO up
to NNLO with respect to the hard pomeron behavior for the parton
distributions at low-$x$. So in the next section we try to do this
comparison for the distribution functions using available results
at NNLO. Eqs.16, 23 and 31-34 are our results for connection
between the gluon distribution and singlet structure function at
small-$x$ by using the Laplace-transform and the hard-Pomeron
behavior (LO up to NLO) respectively. Therefore we show that Eq.1
is not generally true and its validity crucially depends on the
splitting functions and
the initial conditions.\\
 \subsection{3.Results and Discussions}
In order to show our results we computed the gluon distribution
function on the l.h.s of formulas (16, 23 and 31) at small-$x$
with respect to the initial conditions according to the Block
distribution [9-12,39-40]. This distribution represent the
spectrum of possible behavior of the proton structure function and
gluon distribution in the region $x>0.00001$ and $0.11\leq Q^{2}
\leq
1200 GeV^{2}$.\\
We begin by illustration the use of the analytical expression in
Eq.16 to derive $G(x,Q^{2})$ from $F_{2}^{\gamma p}(x,Q^{2})$ in
the case of Refs.[9-12,39-40]. We take the published initial
distributions as our basic input at $Q_{0}^{2}=1 GeV^{2}$, and use
this distribution to calculate the proton structure function
needed in $\mathcal{FF}(x)$. Then, we solve Eq.16 for the gluon
distribution by this $F_{2}^{\gamma p}(x,Q^{2})$ and compared the
results with the published gluon distributions.  In Fig.1 we show
the LO $x$-space results for the gluon distribution for two
representative values of $Q^{2}$. The curves are the published
Block [9-12] gluon distribution, Donnachie and Landshoff (DL) [18-21], GRV-HO [41-42] and GJR parameterization [43].\\
In Fig.2 we present results for the gluon distribution at LO and
NLO using the hard-Pomeron behavior for the gluon distribution
function. This seems to indicate that the gluon distribution is
dominated
 at small-$x$ by  hard-Pomeron exchange. This powerful approach to the
 small-$x$ data for $G(x,Q^{2})$  extends the Regge
 phenomenology that is so successful for hadronic processes. The
 hard intercept is $\delta=0.437$ and we choose $\Lambda$ such
 that $\alpha_{s}(M_{Z}^{2})=0.116$, this gives
 $\Lambda_{n_{f}=4}^{NLO}=400 MeV$ [19-21]. We compared our results by
 published
Block [9-12] gluon distribution, Donnachie and Landshoff (DL)
[18-21], GRV-HO [41-42] and GJR parameterization [43]. As can be
seen, the values of the gluon distribution function increase as
$x$ decreases. This is because the hard-Pomeron exchange defined
by DL model is expected to hold in the small-$x$ limit. Comparing
our results in Figs.1 and 2 with other results indicates that our
global solution (Eq.16) and hard-Pomeron solution (Eqs.23, 31) at
the simplest case are compatible with other phenomenological
models and this is the reason why the approximate
relation (1) is not valid at small-$x$.\\
In order to compare our results with the experimental data, using
Eq.34 for the evolution of the proton structure function with
respect to the gluon distribution function. We show a plot of the
proton structure function in Fig.3 for values of $Q^{2}=8.5~
GeV^{2}$ and $20~ GeV^{2}$, compared to the values measured by the
H1 collaboration
 [45-46] and a QCD fit based on ZEUS data [39-40]. For each $Q^{2}$,
 there is a cross-over point for both the curves where both the
 predictions are numerically equal. As we wants to have a good
 comparison between our results and others, we have to include the
 singlet distribution functions in DGLAP evolution equations.
 However, as there is yet no such simple relation between the
 singlet structure function and gluon distribution at LO up to
 NNLO, we rather appeal to the numerical results of Ref.[38]. In Fig.4 we show
 the ratio $\frac{G(x,Q^{2})}{F^{S}_{2}(x,Q^{2})}$ at $Q^{2}=20
 GeV^{2}$. In this figure we show that this ratio is hardly
 negligible. In order to have more accurate solution for the
 proton structure function, we need to a best global fit for this
 ratio, using NNLO analysis data in Fig.4. We compared our results
 for the proton structure function at NNLO with H1 data [45-46] and
 GJR parameterization [43] and also the gluon distribution function at $Q^{2}=20 GeV^{2}$ in
 Fig.5. It is clear from this figure for $F_{2}$ and $G$ that our
 results, at NNLO analysis and considering of the singlet parton
 distribution, are comparable with other results.\\
In conclusion, the simple relation (1) between the gluon
distribution and singlet structure function is not generally valid
at small-$x$. We show that the gluon distribution can be estimated
with respect to the splitting functions and initial conditions in
a general model by using a Laplace-transform method and
hard-Pomeron model at LO and NLO. Therefore our results at
simplest approach lead to different results from those at
Refs.[4-8]. Further, we need the singlet structure function at the
DGLAP evolution equations for the numerical relation between the
gluon distribution and single structure function. Moreover we
proposed one general numerical approach
at NNLO for this connection and conclude that this numerical approach is agreeing with others results.\\

$\bf{Acknowledgements}$   Author would like to thank the anonymous
referee of the paper for his/her careful reading of the
manuscript and for the productive discussions.\\

\subsection{Appendix}
The NLO splitting function for the singlet structure function is
as follows
\begin{eqnarray}
p^{2}_{qg}&=&2C_{F}N_{f}T_{R}{\{}4-9x-(1-4x)\ln{x}-(1-2x)\ln^{2}x\nonumber\\
&&+4\ln(1-x)+[2\ln^{2}(\frac{1-x}{x})-4\ln(\frac{1-x}{x})-\frac{2}{3}\pi^{2}\nonumber\\
&&+10]P_{qg}(x){\}}+2C_{A}N_{f}T_{R}{\{}\frac{182}{9}+\frac{14}{9}x+\frac{40}{9x}\nonumber\\
&&+(\frac{136}{3}x-\frac{38}{3}){\ln}x-4\ln(1-x)-(2+8x)\ln\ln^{2}x\nonumber\\
&&+2P_{qg}(-x)S_{2}(x)+[-\ln^{2}x+\frac{44}{3}{\ln}x-2\ln^{2}(1-x)\nonumber\\
&&+4\ln(1-x)+\frac{\pi^{2}}{3}-\frac{218}{9}]P_{qg}(x){\}}
\end{eqnarray}
where $P_{qg}(x)=x^{2}+(1-x)^{2}$ and
$S_{2}(x)={\int_{\frac{x}{1+x}}^{\frac{1}{1+x}}\frac{dz}{z}\ln(\frac{1-z}{z})}$.
The small-$x$ limit of the NLO splitting function for the
evolution of the singlet quark is then [44]
\begin{equation}
p^{2}_{qg}{\longrightarrow}\frac{\alpha_{s}}{4\pi}\frac{80C_{A}N_{f}T_{R}}{9x}.
\end{equation}
where the casimir operators of colour SU(3) are defined as $
C_{A}=3, ~C_{F}=\frac{4}{3}$ and $T_{R}=\frac{1}{2}$.\\
\begin{figure}
\includegraphics[width=0.5\textwidth]{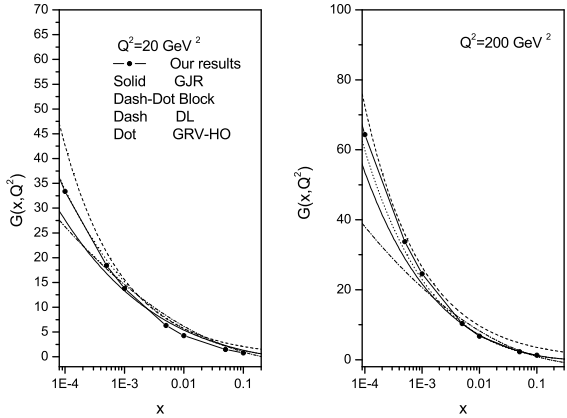}
\caption{The gluon distribution, $G(x,Q^{2})$, for $Q^{2}=20$ and
$200 GeV^{2}$ by using a Laplace-transform. Our results compared
to the Block model [9-12], DL model [18-21], GRV-HO
parameterization [41-42] and GJR parameterization
[43].}\label{Fig1}
\end{figure}
\begin{figure}
\includegraphics[width=0.5\textwidth]{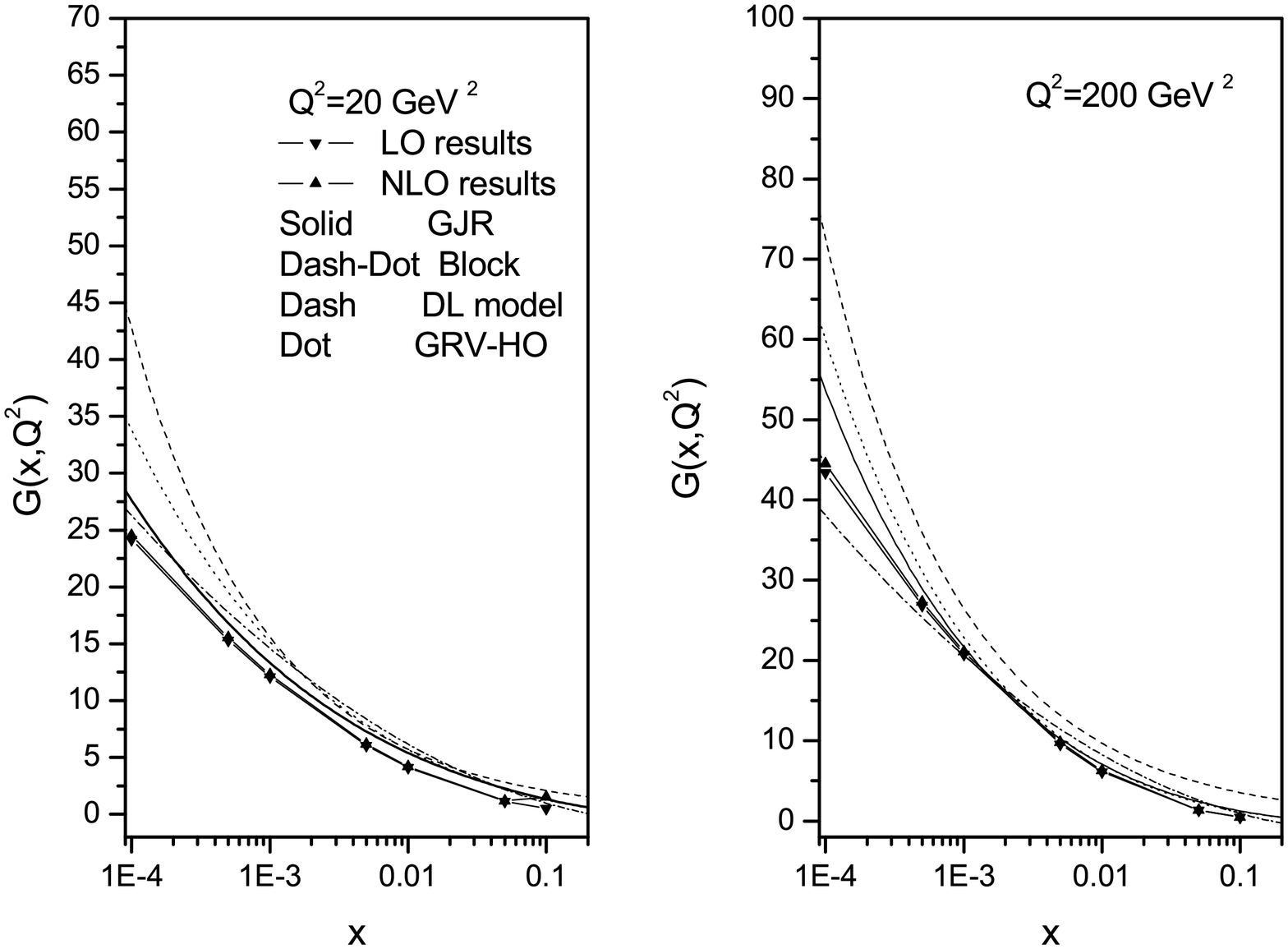}
\caption{The same as Fig.1 but for a hard-Pomeron behavior at LO
and NLO.}\label{Fig2}
\end{figure}
\begin{figure}
\includegraphics[width=0.5\textwidth]{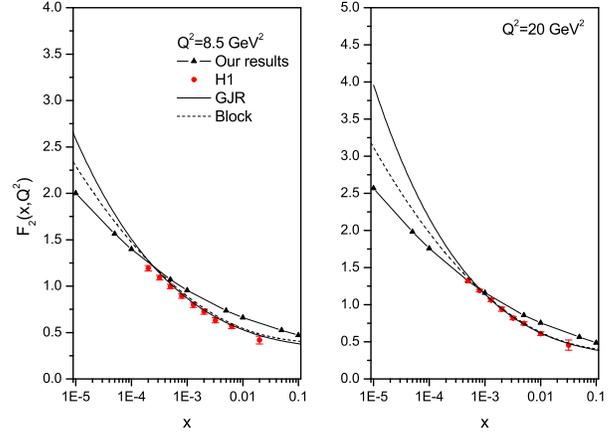}
\caption{The proton structure function, $F_{2}(x,Q^{2})$, for
$Q^{2}=8.5$ and $20 GeV^{2}$ with respect to a hard-pomeron
behavior. Our results compared to the Block model [9-12], GJR
parameterization [43] and H1 data [45-46].}\label{Fig3}
\end{figure}
\begin{figure}
\includegraphics[width=0.5\textwidth]{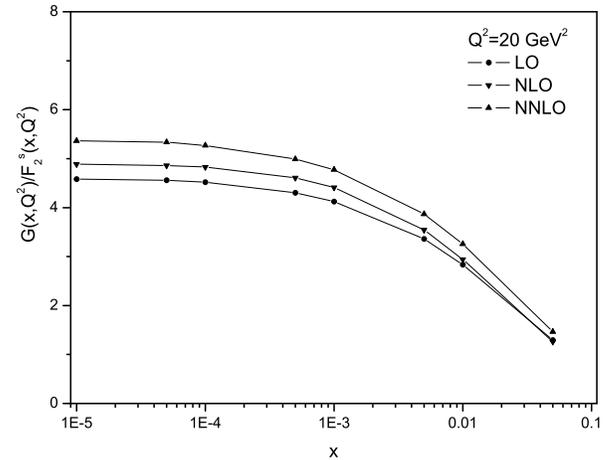}
\caption{The ratio $\frac{G(x,Q^{2})}{F^{S}_{2}(x,Q^{2})}$ at LO
up to NNLO when we consider the complete form of the DGLAP
evolution equations. }\label{Fig4}
\end{figure}
\begin{figure}
\includegraphics[width=0.5\textwidth]{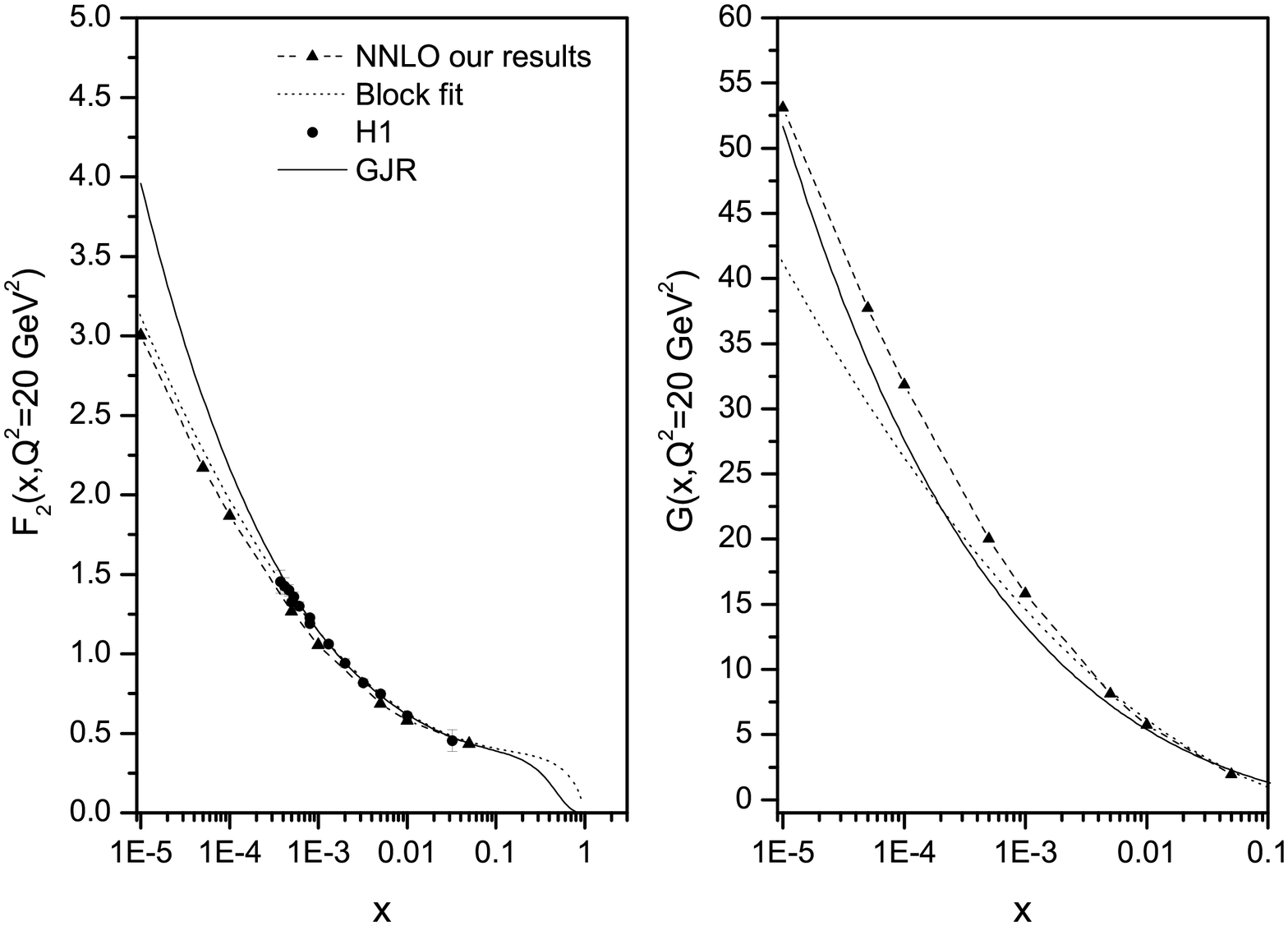}
\caption{Numerical estimates at $Q^{2}=20~GeV^{2}$, for the proton
structure function and gluon distribution function at NNLO
analysis and at the complete form of the DGLAP evolution
equations, compared with other results. }\label{Fig5}
\end{figure}

\section{References}

1. Yu.L.Dokshitzer, Sov.Phys.JETP {\textbf{46}}, 641(1977).\\
2. G.Altarelli and G.Parisi, Nucl.Phys.B \textbf{126},
298(1977).\\
3. V.N.Gribov and L.N.Lipatov, Sov.J.Nucl.Phys. \textbf{15},
438(1972).\\
4. J.K.Sarma and B.Das, Phys.Lett.B {\bf304}, 323(1993).\\
5. J.K.Sarma, D.K.Choudhury and G.K.Medhi, Phys.Lett.B {\bf403},
139(1997).\\
6. R.Baishya and J.K.Sarma, Phys.Rev.D {\bf74}, 107702(2006).\\
7. R.Baishya, U.Jamil and J.K.Sarma, Phys.Rev.D {\bf79},
034030(2009).\\
8. M.Devee, R.Baishya and J.K.Sarma, Eur.Phys.J.C {\bf72}, 2036(2012).\\
9. M.M.Block, L.Durand, D.W.McKay, Phys.Rev.D\textbf{77},
094003(2008).\\
10. M.M.Block, L.Durand, D.W.McKay, Phys.Rev.D\textbf{79},
014031(2009).\\
11. M.M.Block, Eur.Phys.J.C\textbf{65}, 1(2010).\\
12. M.M.Block, L.Durand, P.Ha and
D.W.McKay, Phys.Rev.D\textbf{83}, 054009(2011).\\
13. G.R.Boroun, JETP\textbf{106}, 700(2008).\\
14. G.R.Boroun and B.Rezaei, Eur.Phys.J.C\textbf{72},
2221(2012).\\
15. B.Rezaei and G.R.Boroun, JETP\textbf{112}, 381(2011).\\
16. A.V.Kotikov, Phys.Rev.D {\bf49}, 5746(1994).\\
17. A.V.Kotikov and G.Parente, Phys.Lett.B {\bf379}, 195(1996).\\
18. A.Donnachie and P.V.Landshoff, Phys.Lett.B {\bf437}, 408(1998 ).\\
19. A.Donnachie and P.V.Landshoff, Phys.Lett.B {\bf550}, 160(2002 ).\\
20. P.V.Landshoff, arXiv:hep-ph/0203084 (2002).\\
21. P.V.Landshoff, arXiv:hep-ph/0209364 (2002).\\
22. K.Golec-Biernat, J.Phys.G {\bf28}, 1057(2002).\\
23. K.Golec-Biernat, Acta Phys.Pol.B {\bf35}, 3103(2004).\\
24. K.Golec-Biernat, Acta Phys.Pol.B {\bf33}, 2771(2002).\\
25. A.M.Stato, K.Golec-Biernat and J.Kwiecinski,
Phys.Rev.Lett.{\bf86}, 596(2001).\\
26. E.Iancu, K.Itakura and S.Munier, Phys.Lett.B {\bf590},
199(2004).\\
27.  E.Iancu, K.Itakura and S.Munier, Nucl.Phys.A {\bf708},
327(2002).\\
28. L. McLerran and R. Venugopalan, Phys.Rev.D\textbf{49},
2233(1994).\\
29. E. Iancu, A. Leonidov and L. McLerran,
Nucl.Phys.A\textbf{692}, 583(2001).\\
30.  E. Iancu, A. Leonidov and L. McLerran,
Phys.Lett.B\textbf{510},
133(2001).\\
31. E. Ferreiro, E. Iancu, A. Leonidov and L. McLerran, Nucl.Phys.A\textbf{703}, 489(2002).\\
32. M.Praszalowicz and T.Stebel, arXiv:hep-ph/1302.4227(2013).\\
33. G.Beuf and D.Royon, arXiv:hep-ph/0810.5082(2008).\\
34. M.Praszalowicz, arXiv:hep-ph/1304.1867(2013).\\
35. J.R.Forshaw and G.Shaw, JHEP{\bf12}, 052(2004).\\
36. J.Kwiecinski and A.M.Stasto, Phys.Rev.D.{\bf66},
014013(2002).\\
37.  F.Caola and S.Forte, Phys.Rev.Lett.{\bf101}, 022001(2008).\\
38. G.R.Boroun and B.Rezaei, Eur.Phys.J.C\textbf{73},
2412(2013).\\
39. M.M.Block, E.L.Berger, C-I Tan, Phys.Rev.Lett\textbf{97},
252003(2006).\\
40. E.L.Berger, M.M.Block, C-I Tan,
Phys.Rev.Lett\textbf{98}, 242001(2007).\\
41. M.Gluk, E.Reya and A.Voget, Z.Phys.C\textbf{67}, 433(1995).\\
42. M.Gluk, E.Reya and A.Voget, Eur.Phys.J.C\textbf{5}, 461(1998).\\
43. M. Gluck, P. Jimenez-Delgado, E. Reya, Eur.Phys.J.C
{\bf53}, 355({2008}).\\
44. R.K.Ellis , W.J.Stirling and B.R.Webber, \textit{QCD and
Collider Physics}(Cambridge University
Press)1996.\\
45. $H1$ Collab., C.Adloff \textit{et al}., Eur.Phys.J.C \textbf{21}, 33(2001).\\
46. $H1$ Collab., C.Adloff \textit{et al}., Eur.Phys.J.C \textbf{71}, 1579(2011).\\

\end{document}